# Chronic Kidney Disease of Unknown Aetiolgy (CKDu) – the search for causes and the impact of its politicization.

By

M. W. C. Dharna-wardana,

National Research Council of Canada, Ottawa, Canada, K1A 0R6,
and
Dept. de Physique, Univsersité de Montreal, Montréal, Canada, H3C 3J7

**Abstract**. Kidney disease of unknown aetiology (CKDu) has been identified in many countries extending from MesoAmerica and Egypt, to South-east Asia and China. Although CKDu has been linked by various authors to farming, it is an artifact of treating multi-modal disease distributions as unimodal. There is NO correlation of CKDu with agriculture since affected farming villages are often surrounded by other farming villages free of CKDu. Initial studies looked for a correlation of CKDu with toxic heavy metal residues of arsenic, cadmium etc., or herbicides like glyphosate that may be present in the environment, as the causative factors. There is now considerable consensus that their concentrations are below danger thresholds, be it in Mesoamerica or south-east Asia. The conceptual basis of a search for etiology within a systems approach is discussed, and attempts to name the disease to bias the identification of its etiology are reviewed. Current research has narrowed down the etiology to geochemical electrolytic contaminants like fluorides and ionic components in hard water, nanosilica (found in water as well as in the air), as well as renal toxins similar to indoxyl sulphates that may arise from interactions of ions with humic acids contained in aqueous organic matter. However, while agrochemical toxins are increasingly considered less relevant to the etiology of CKDu, it has become a firm public belief. In Sri Lanka this has spawned ideology-based agricultural policies for partial and complete banning of agrochemicals (2014-2021), followed by some back-tracking, disrupting the economy and the food supply. A farmer's uprising in 2022 was spawned by poor harvests. It triggered a larger popular uprising that led to the collapse of a government wedded to romanticized eco-extremist agricultural policies in a country already facing difficulties in the wake of Covid and Ukraine.
[A shorter, peer-reviewed version of this article appears as Chapter 17 of *Medical Geology, Enroute to One Health*, Eds, Prasad and Vithanage, Wiley (2023)]

**Introduction**. Epidemics of chronic kidney diseases of unknown aetiology (CKDu) have been identified in several parts of the world since the 1990s, e.g., as reviewed in Hettithanthiri et al (2021), or Weaver et al (2015). Hints of this unusual form of kidney disease existed as early as the 1970s (Wessling et al 2015). The disease differs from the more familiar form of chronic kidney disease (CKD) in that CKDu is a 'silent disease' that progresses without the usual signs of hypertension, diabetes and glomerular disease found even in the early stages of standard CKD (Jayatilleke 2013). Epidemics of CKDu having a similar lack of early symptoms, and usually leading to tubulo-interstitial damage in the kidneys (Nanayakkara et al 2012, Wijkström et al 2020) have been observed in many parts of the world. Thus reports on the disease are available from Sri Lanka (Jayatilleke et al 2013, Dissanayke et al 2005), Mesoamerica (Correa-Rotter et al 2014, Sanchez Polo et al 2020), Mexico (Trabanino et al 2005, Aguilar-Ramirez et al 2021), India (Andra Predesh, Rajkapur et al 2012, Gowishankar et al 2020; Tondaimandalam in Tamil Nadu, Naraisinghpur and Badamba in Odisha, Sahil et al 2021), Thailand (mainly North-East, Aekplakorn et al 2021), and El Minya Governorate in Egypt (El Minshway et al 2011) etc. Although there are clear similarities in the manifestation of the disease in different geographic regions, the etiological causes may well be different, and a wide range of causative factors has been suggested. Researchers into Mesoamerican CKDu have emphasized heat stress while also considering the implications of agrochemical use. South-Asian CKDu research initially emphasized agrochemical toxins, but has increasingly turned to causative factors found in drinking water. Agrochemical residues found in the soil and water in Mesoamerican (Gonzales-Quiroz et al 2018) and south-Asian (Hettithanthiti et al 2021) endemic areas have consistently proven to be well below the accepted danger thresholds. Furthermore, agrochemical use is ubiquitous, while the disease is confined to specific geographic areas.

We will not review the disease since excellent reviews are available, e.g., in Hettithanthiri et al (2021). We focus on conceptual and methodological issues bearing on the etiology of the disease, premature implication of etiology *via* biased nomenclature, the politicization of the study of the etiology of the disease and consequences to public policy.

**Conceptual Issues.**
The etiology of non-communicable chronic diseases (NCCDs) is notoriously difficult since many causative factors associated with life styles, genetics, infections, environmental factors etc., and come into play. No consensus exists even for well-researched NCCDs (e.g., Okuyama et al 2018), e.g., even on the "Consensus Statement" of the European Atherosclerosis Society Consensus Panel (Ference et al 2017). Nevertheless, the challenge to the public-health scientists is precisely to narrow down the causes to a minimum number, given an accepted template of life styles, environmental and genetic factors in a given population. Etiology requires going beyond mere associations of the disease with some factor and presenting a causative relationship even when it may be without a clear physiological mechanism.

Various internationally accepted specifications of maximum acceptable daily intake (ADI) levels (e.g., as given in the Cordex Alimentarius) are for *individual* toxins. Their interactions need to be taken into account in practical applications. Thus the toxic effects of traces of cadmium may be largely mitigated by the co-presence of other ions like zinc or selenium (Chaney 2012, Dharma-wardana 2018) leading to the safe consumption of shell fish, sunflower oil, potatoes, rice etc. as practiced in many parts of Europe.

One of the simplest approaches to the conceptual problem of sorting out associations and causations is proposed by the Bradford Hill criteria (Hill 1965). Hill asked, "In what circumstances can [we] pass from [an] observed association to a verdict of causation?" He proposed nine "aspects of association" for evaluating traditional epidemiological data. These aspects are now known as the Bradford Hill (BH) Criteria.

The nine "associations" that Hill discussed (strength of association, consistency, specificity, temporality, biological gradient, plausibility, coherence, experiment, and analogy) identify causative agents that should finally be pinned down in a physiological model. Surprisingly, while over three dozen causative factors for CKDu have been proposed, these proposals have not even used the BH criteria to eliminate the more unlikely hypotheses, and instead the "multi-factorial" character of CKDu has been emphasized in a superficial way.

Even the use of the BH criteria would have been enough to avoid naive claims of increasing-glyphosate use being causatively associated with autism or with various NCCDs. Such claims are mostly found on the internet or in publications that have appeared in predatory journals, and can have a disproportionate effect on public policy.

The modern approach to causative factors is strongly transclinical since advances in chemical analytical techniques, molecular biology, molecular toxicology, information technology, mathematical modeling as well as the design of experiments using cell lines and laboratory animals have provided a deeper and more complex understanding of systems-based etiology. So the inference of "association to causation" needs the support of a physico-mathematical model unless a clear cause stands out.

Once likely causative factors (agents) *f1, f2, f3, ...* , have been identified, their interactions can be represented by a matrix of interactions (e.g., see the appendix to Dharma-wardana et al 2015) that can be diagonalized (at least approximately) to carry out a factor analysis (e.g., see Paranagama 2012). Then *f1, f2, f3, ...*, are replaced by their linear combinations giving new combined factors (eigenvectors) *F1, F2, F3, ...*, and show how the agents intermix synergistically or antagonistically. A synergy is complete "in-phase" action, while an antagonism is complete "out-of-phase" action, while intermediate phase angles between 0 and π may occur if phase shifts (time delay in actions) can occur. An organigram can be used to present the interactions as a complex interacting system. This provides a means of visualizing not only the interactions, but also the non-linear effects of negative and positive feedback loops in the interactive system (Jayasinghe et al 2020, Dharma-wardana 2020).

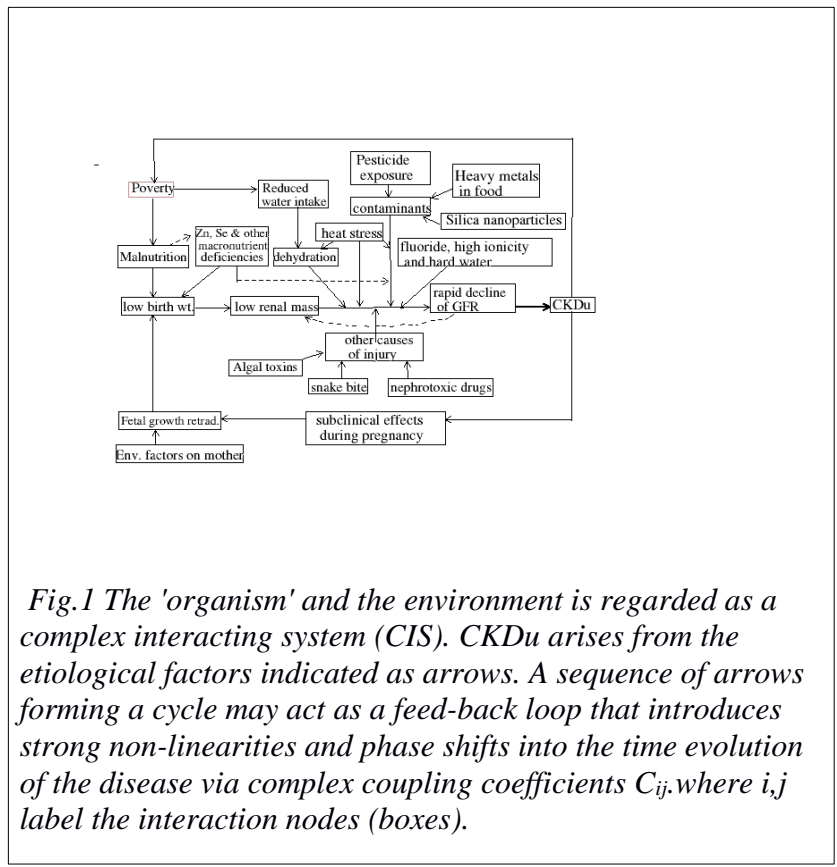

*Fig.1 The 'organism' and the environment is regarded as a complex interacting system (CIS). CKDu arises from the etiological factors indicated as arrows. A sequence of arrows forming a cycle may act as a feed-back loop that introduces strong non-linearities and phase shifts into the time evolution of the disease via complex coupling coefficients $C_{ij}$.where i,j label the interaction nodes (boxes).*

Such an organigram (Fig. 1) is a schematized *complex interactive system* (CIS) containing the organism in its environment. The feed-back processes introduce non-linear causative evolution of the pathology. Unfortunately, the strengths and weights (also known as matrix elements or coupling coefficients $C_{ij}$) that are associated with the arrows that connect different nodes *i,j* (boxes in Fig. 1) are not known, but they are needed as a function of time to make any use of the rich mathematical theory of complex systems.

The Complex Interacting System may be dominated by a major causative factor (i.e., a dominant arrow in the diagram at a given time). In less clear-cut cases, these diagrams depict the various inter-relations among causative factors in a specific environment. The causative factor triggering the disease can differ from that propelling the evolution of the disease in each location. For instance, intake of silica nanoparticles may be highly relevant for CKDu among Nicaraguan workers exposed to aerosol ash produced in burning sugarcane. The burning of paddy husks may also generate silica nanoparticles; however, paddy-husk burning is practiced not only in the endemic region (villages scattered in Andra Predesh, or in the endemic areas of Sri Lanka), but also in other villages where CKDu is not found. However, elevated levels of silica in drinking-water wells (Nikagolla 2020) in endemic regions is relevant and needs further study. Similarly, the mechanism based on fluoride and hard water relevant to the dry zone villages in Sri Lanka (Dissanayake et al 2005, Wasana et al 2018, Dharma-wardana 2018, Balasooriya 2018) may be irrelevant to Nicaragua, while possibly relevant to Andra Pradesh too.

Risk factors like hypertension, diabetes etc., appear along with the onset of the disease in conventional CKD. Uremic toxins like indoxyl suphate and p-cresyl sulphate have been considered in the etiology of CKD (e.g., Wang and Zhao 2018). Makehelwala et al (2020) have studied dissolved organic carbon interactions and uremic toxins formed by complexation with $Ca^{2+}$ and $SO_4^{2-}$ ions in endemic areas. This merits further study, although such uremic toxins should cause conventional CKD rather than CKDu. McDonough et al (2020) have examined the water chemistry and the microbiome of household wells in CKDu-endemic Medawachchiya, Sri Lanka. They report high levels of fluoride, magnesium, sodium, chloride and calcium in some samples; they also frequently find cyanotoxin-producing Microcystis. Why such toxins should cause CKDu rather than CKD is not clarified. Furthermore, not enough information is contained in these initial explorations to apply even the BH criteria for etiology, thus limiting the impact of such studies.

The *exposome concept* (Wild 2005, Miller et al 2014) is an ambitious attempt to provide a more accurate environmental-impact assessments on human health using large numbers of measurements of all relevant health and environmental parameters over the life time of test individuals. Thus the EXPOsOMICS project (exposmics 2013) aims to assess environmental exposures, focusing on air pollution and water contaminants, and link them to bodily biochemical and molecular changes.

These very expensive data-intensive methods may be relevant to further studies on well-studied NCCDs like diabetes and obesity. At a simpler level, Anand et al (2019) have discussed molecular and genetic methodologies for the investigation of the CKDu epidemiology and conclude that "research has been challenging because of political circumstances, the marginalized nature of populations afflicted, and the scarcity of personnel and funding". Given these bottlenecks for CKDu research, even basic statistical data on disease prevalence are sadly lacking. Thus, in Sri Lanka it is left to the rural patient to maintain his/her health records (Nanayakkara 2012) and hence there can be no reliable "official data set" for the epidemic in Sri Lanka. So no proper disease statistics exist, although researchers have drawn maps giving the impression of the availability of good data. Ranasinghe et al (2019) in their survey point out widespread discrepancies in quoted data in studying some 30,000 patients who included *both* CKD and CKDu. What is reliably available is mostly estimates from limited surveys done during etiological studies, e.g., in the study of CKDu sponsored by the Sri Lankan National Science Foundation (NSF) and the World Health organization (Jayatilleke et al 2013). The same situation holds for many developing countries afflicted with CKDu.

**Misuse of disease nomenclature.**
The name "chronic kidney disease of unknown etiology" (CKDu) has been in use since 2006. Its use became more prevalent after its adoption by the International Society of Nephrology via the *International Consortium of Collaborators on Chronic Kidney Disease of Unknown Etiology in 2018*. A number of other names (e.g., Mesoamerican nephropathy, Balkan endemic nephropathy, Sri Lankan nephropathy, Uddanam nephropathy (Andra Pradesh) had also been used, especially to emphasize geographic origins.

Another more questionable type of nomenclature embeds an assumed etiology or makes a link to a specific occupation (e.g., agriculture) quite prematurely. In our view, the use of such names, e.g., "chronic interstitial nephritis in agricultural communities" (Jayasumana et al 2016), "kidney disease of unknown cause in agricultural laborers" (subramanium 2017), or directly linking etiology to agrochemicals, as in the appellation "chronic agrochemical nephropathy" (Jayasinghe 2014), are such examples of premature naming. The latter was proposed in 2014 although several studies (e.g., Nanayakkara et al 2012) of CKDu in Sri Lanka had begun to indicate that agrochemical toxins in the endemic areas were well below the danger thresholds. This was a consequence of the reasonable worry about excessive use of agrochemicals that continues to be a part of public thinking. Corresponding studies in Mesoamerica had also suggested that the etiology of the disease cannot be pinned clearly to agrochemicals. These premature linking of CKDu with agrochemicals spawned public fear and disastrous socio-political consequences discussed in the last section of this study.

**Is CKDu a disease associated with agricultural communities?**
The endemic regions are mainly in tropical rural areas where the residents live by small-scale agriculture (e.g., paddy farming in Sri Lanka or in India). In contrast, Mesoamerican CKDu is found among sugar cane workers employed by large agri-business. Some authors have renamed the disease as "agricultural nephropathy" or "chronic interstitial nephritis in agricultural communities (CINAC)", implying and imposing a well-established etiology by the *act of naming* by itself. Ververt et al (2020) have even claimed that dysmorphic lysomes detected using electron microscopy of tissues from "CINAC" patients directly point to an agricultural toxin, while this has been contested by Wijkstrom et al (2020).

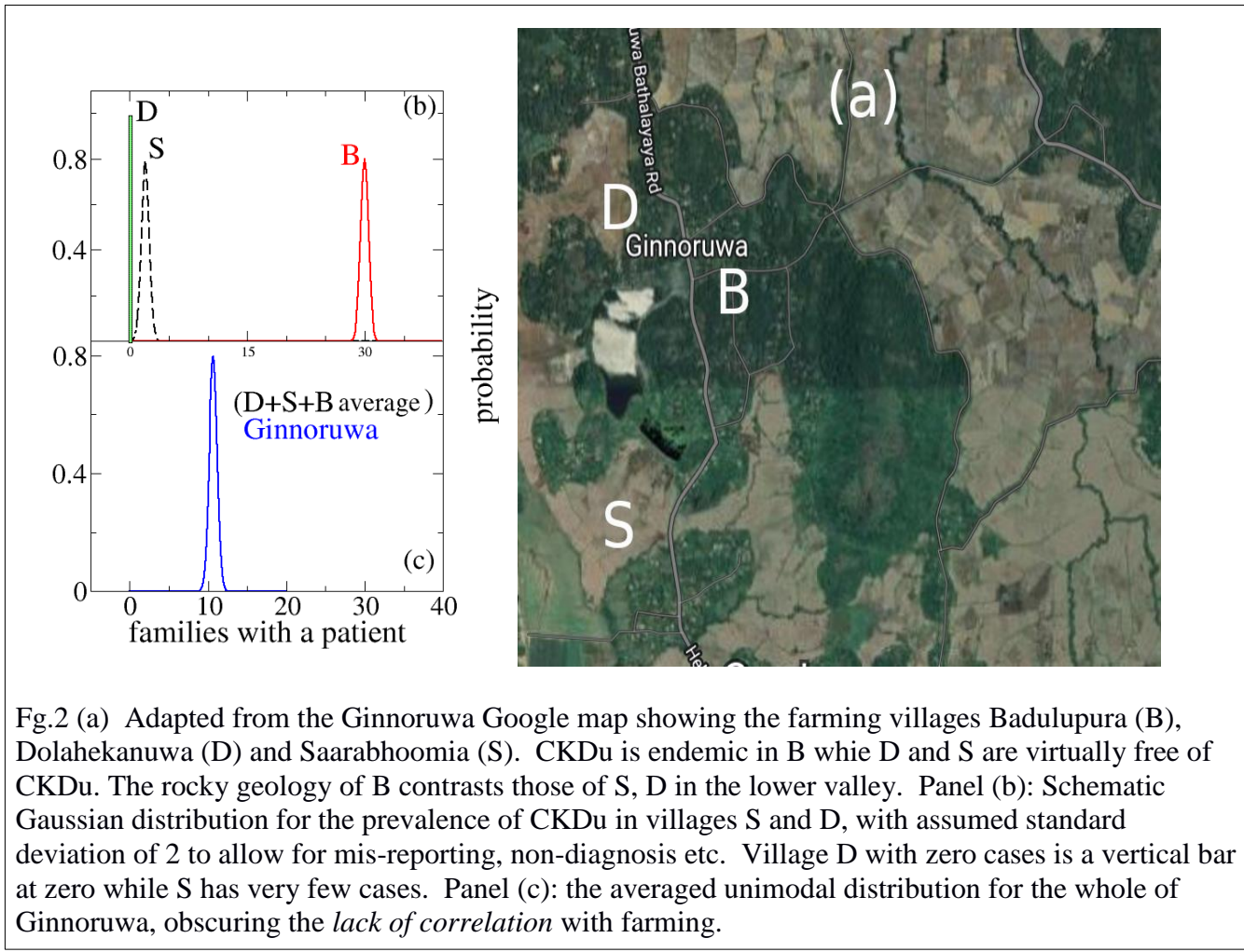

Fg.2 (a) Adapted from the Ginnoruwa Google map showing the farming villages Badulupura (B), Dolahekanuwa (D) and Saarabhoomia (S). CKDu is endemic in B whie D and S are virtually free of CKDu. The rocky geology of B contrasts those of S, D in the lower valley. Panel (b): Schematic Gaussian distribution for the prevalence of CKDu in villages S and D, with assumed standard deviation of 2 to allow for mis-reporting, non-diagnosis etc. Village D with zero cases is a vertical bar at zero while S has very few cases. Panel (c): the averaged unimodal distribution for the whole of Ginnoruwa, obscuring the *lack of correlation* with farming.

In reality, this claimed association between CKDu and agriculture is a chimera of bad statistics. That is, there is *not even an association,* leave aside a causative implication between CKDu and agriculture. A whole region e.g., the North Central Province (NCP) of Sri Lanka, or Andra Predesh, India, where agriculture is the main occupation is superficially surveyed, and averaged-out statistics are given for the number of people stricken by CKDu. Naturally, CKDu patients are *ipso facto* farmers because most people in these regions *are* farmers. Such averages are meaningful only if the underlying statistical distribution is unimodal. We show below that the distribution in not unimodal.

A more microscopic view of the distribution of CKDu shows that CKDu-striken farming villages have other adjacent farming villages with essentially *no incidence* of CKDu. The Ginnoruwa region of the NCP of Sri Lanka is a well-studied endemic area (Balasooriya et al 2018) consisting of a cluster of three adjoining villages (Fig. 2), namely Badulupura (B), Dolahekanuwa (D) and Sarabhoomiya (S).

These villages were settled during 1980s for agriculture via the Mahaweli-River development project, each having about 100 families. At least 30% of the families in B have a CKDu patient. The adjacent villages, S and D have low (<2%) or zero records of families with CKDu patients, even though they lead similar lives, eat nearly identical diets, have close family links and cultivate paddy in a common low-lying area.

Very few patients were reported in the S community and even they had come from the adjacent village B and settled in S, implying that the disease was triggered during their sojourn in B. Some 70% of all affected CKDu patients were male, but this percentage is sensitive to how the e-GFR criterion for CKDu is used. For instance, the NSF-WHO sponsored study (Jayatilleke et al 2013) of the whole NCP claimed that "the overall prevalence of CKDu was 15.3% with a higher prevalence in females (16.8%) than males (13.3%). More severe grades of CKDu were seen more frequently in males (grade 3: males vs. females=19.9 vs 5.3%, grade 4; males 16.1% vs females 3.8%. In both sexes prevalence was higher with increasing age". Jayalal et al (2019) also studied Sri Lankan CKDu and concluded that a higher percentage of females were affected. In contrast, most researchers of CKDu in Sri Lanka and elsewhere have concluded that the disease is more prevalent in males. The differing conclusions regarding gender bias can be traced to (i) uncertainties in identifying patients in the asymptomatic early stages (grades 1, 2) in using e-GFR data, (ii) researchers relying on data based on reports by rural people rather than on getting data by conducting actual field tests. It should be emphasized that doctors don't maintain records within the Sri Lankan health system, it being left to the patients themselves (Nanayakkara et al 2012).

Balasooriya et al (2018) state that nearly 98% of the population in the area (Fig. 2) use groundwater as the primary source for drinking water. Rice is the main food crop grown in the area and the population mostly consumes rice as staple food obtained from their own paddy lands. No noticeable differences between socio-economic status and food consumption behavior were observed between CKDu and non-CKDu families in the study region. No ethnic differences were present in the patient population. They were settlers who came to the Ginnoruwa from the Badulla area, south east Sri Lanka.

Badulupura (B) has a rocky geology, Fig. 2(a), while Fig. 2(b) shows the distinct disease distributions for the three villages. The distribution marked B, D, S schematizes the 30%, and zero or ~2% cases. Clearly, there is NO correlation between CKDu and farming; although the residents in D, S are also farmers, they show no CKDu. So CKDu is not correlated with farming unless a unimodal average distribution Fig.2 (c) is incorrectly used.

Thus, associating CKDu with agricultural activities is a consequence of using averages over-arching multi-modal distributions. It is very likely that this holds true for CKDu in Andra Predesh and other endemic areas too. Hence names like "agricultural nephropathy" are misnomers for this disease. There is NO real association of CKDu with agricultural activity as seen in Ginnoruwa and in other examples from Sri Lanka.

**Is CKDu a disease associated with the use of agrochemicals**?
The hypothesis that agrochemicals may be the cause of CKDu is a very natural one, even for those who view the impact of the "green revolution" positively. Since the CKDu areas are rural and pastoral, the focus has been on the presence of toxic impurities in fertilizers, and the impact of excessive use, or indeed *any use* of pesticides. It is claimed by a small but powerful elite segment of society that food grown using fertilizers and pesticides is inherently unsafe for heath, and that one must become "sync with nature". That CKDu is caused by toxins in agrochemicals (fertilizers and pesticides) is a straight forward step for those who begin from such a "naturalistic" view point.

The spokesman for the Government Medical Officer's Association of Sri Lanka, a medical doctor named Anuruddha Padeniya (Padeniya 2020) had released an influential U-tube video claiming that ancient Sri Lankans, fed on an indigenous diet had lived to 140 years of age, and that the introduction of agrochemicals had led to a "poisoned diet". Padeniya also designated CKDu as an agrochemical nephropathy and indicted the use of agrochemicals for causing the CKDu epidemic. Padeniya's opinions were very influential among lawmakers who quoted him in parliament, although many distinguished scientists made futile efforts to set the record right (e.g., Pethiyagoda 2021).

Dr. Ranil Senanayake, an eco-activisist and militant supporter of organic farming had claimed that the incidence of non-comunicable chronic diseases (NCCDs) in Sri Lanka had grown exponentially since the 1970s. These views were taken up by popular politicians who called for a return to traditional agriculture "free of toxins". However, Dr. Senanayake's own data were found to show a mere linear growth of NCDs consistent with the gradual rise in life expectance and longevity in Sri Lanka that had occurred during the period. This data and the associated discussion may be found under BFBF (2021) and led to public discussions in newspapers (Dharma-wardana 2022a).

We first examine the claim that traces of heavy-metals and metalloids like arsenic present in fertilizers added to soils constitute a possible cause of toxins in the food chain.

*Impurities in Fertilizers.*
The view that the regular addition of fertilizers containing trace toxic components (e.g., phosphate fertilizers containing Cd, As etc) will eventually increase the concentration of these toxins in the soil and cause chronic disease is widely entrenched since the early work of Schroeder and Balassa (1983). It is incorrect for the most part, as briefly explained below. More extensive discussions, mainly focusing on Cd in soils, water and food are available (Dharma-wardana 2018b, Roberts 2015, Cheney 2012) and can be easily generalized to other toxins.

To lay bare the paradigm of Cd accumulation by fertilizer inputs, we calculate the incremental change in the soil-cadmium concentration ($\Delta C$) on addition of a "contaminated" phosphate fertilizer to the soil.
- Let 25 kg of phosphate fertilizer/ha be applied to a depth of 15 cm in paddy cultivation.
- The corresponding soil volume/ha is $V_s = 15 \times 10^5$ litres of soil, treated with fertilizer.
- Let the density of soil be 1.3 kg/litre, then the total weight of the soil is $19.5 \times 10^5$ kg, i.e., approximately 2 million kg of soil/ha.
- Given current European Union standards, a highly "contaminated" sample of TSP unacceptable in the EU may contain 50 mg of Cd, and 50 mg of As per kg of fertilizer (this is not considered contaminated by US standards).
- Hence 25 kg/ha addition of such a contaminated fertilizer will add only $\Delta C = (50 \times 25$ mg$)/(2,000,000) = 625$ ng per kg of soil.

A nanogram per kg is one part in a trillion. Thus the Cd increment added to the soil via highly contaminated fertilizer application, even if we completely ignore leaching and other removal processes is in parts per trillion! This is utterly negligible compared to the typical ambient concentrations of Cd in agricultural soil (approximately 1-3 mg Cd/kg of soil). So, even if the added cadmium accumulates for a millennium even without any leaching, any removal effects or conversion to bio-unavailable insoluble forms, the effects are quite imperceptible.

Furthermore, if the extensive drainage during monsoonal rains found in CKDu-endemic South-Asian regions is taken into account in addition to the usual leaching processes and removal processes, the effect of the heavy-metal or metalloid toxins added to the soil reduces essentially to zero.

So the proportionately increasing Cd content observed in crops with increasing fertilizer inputs discussed in traditional studies (e.g., Silanpää and Jansson 1992) comes from the naturally existing cadmium in the soil, triggered by chemical changes of the soil pH, and from enhanced microbial activity due to the nourishment provided by fertilizers. In the case of organic fertilizers, their P, N, content is usually very low and quite variable, e.g., 1-3% of phosphate per kg of manure. Hence, to obtain phosphate inputs comparable to those from mineral fertilizers, several tonnes/ha of organic fertilizer are needed where only a few kg/ha of mineral fertilizers suffice. Here too, (i.e., even with organic fertilizer) soil micro-organisms are activated and trigger an increase in cadmium uptake by plants where cadmium fixed in the soil is made bioavailable by bacterial action or chemical action. Similar heavy-metal effects should arise in the use of bio-film bio-fertilizers (BFBF 2021) as well.

The argument that the amount of cadmium or arsenic found in the soil decreases as we dig down, and that they must have come from surface addition of fertilizers has sometimes been used in non-scientific discussions in TV and in other social media. This argument fails to realize that most heavy metals in the soil result from the weathering of rocks during the formation of top soil. Highly industrialized neighborhoods may contain such contamination from mining, smelting, or coal-fired power stations, or even automobile traffic (as with lead) that generate aerial deposition. These are not applicable to the CKDu endemic regions discussed here, although lead deposition near highways from the use of lead containing fuels is a possibility.

We note that most environmental studies in the CKDu endemic locations have reported quite low levels of metal or metalloid toxins. This is established for Sri Lanka in Nanayakkara et al (2012), Levine et al (2016), Balasooriya et al (2019) and in other studies reviewed in, e.g., Hettithanthiri et al (2021). Similar conclusions are found in Mesoamerican studies as well (Korrea-Rotter et al 2014, Gonzales-Quiroz et al 2018, Mazeyra et al 2022); consequently they have addressed causes other than agrochemicals, e.g., heat stress or nanosilica.

We review the case of Ginnoruwa, Sri Lanka. The three villages B, D and S (discussed in the previous section) use water from dug wells for drinking and cooking. The geology of Fig. 2(a) shows that Badulupura wells are *not* connected to the lower water table of the agricultural area but are sourced by regolith aquifer water. Isotopic studies (Wickremarathna et al 2017, Edirisinghe et al 2017 ) have verified this. The safely usable wells (in S, D) are connected with the agricultural water table known to contain negligible agrochemical toxins. The high-risk CKDu wells also contain negligible agrochemical toxins, but they are rich in fluoride and magnesium ions (Balsooriya et al 2019). Fluoride ions present in hard water were proposed to be nephrotoxic from studies on laboratory mice (Wasana et al 2017). Furthermore, theoretical considerations were used to propose that $Mg^{++}$ ions in the hard water and $F^-$ ions act synergistically via $MgF^+$ complex formation in the water (Dharma-wardana 2018). When Mg and F ions are found in significant aqueous concentrations, they are known to be highly active Hofmeister ions in protein denaturing (Dharma-wardana et al, 2015).

These results are consistent with experimental studies on effluent from reverse-osmosis of drinking water (Imbulana et al 2020), and on studies of CKDu in Moneragala, Sri Lanka (Liyanage et al 2022) where higher concentrations of $Mg^{2+}$ & $F^-$ ions were found in CKDu-associated water sources. In contrast, a factor-analysis study (Paranagama etal 2018) claimed stronger correlation between $F^-$ & $Na^+$ than with $F^-$ & $Mg^{++}$ in CKDu-linked water sources. This highlights the difficulty of moving from correlations to causes unless a physiological mechanism (e.g., co-absorption of $MgF^+$ or some other entity as a complex) can also be investigated.

The villages B, D and S together form a tri-modal (or nearly bi-modal) distribution of CKDu incidence. Such multi-modality is probably typical of how healthy villages intersperse among CKDu villages.

Hence engaging in agriculture is not even associated with CKDu, and furthermore, there is no significant presence of elevated levels of agrochemical toxins in the drinking water obtained from dug wells, or rivers, irrigation tanks and canals (Jayasinha et al 2015) of the endemic regions.

**Are pesticide residues implicated?**
Historical use of DDT in the 1950s led to successfully control of Malaria that nearly brought the then health-care system of the country to its knees. The NSF-WHO sponsored study of CKDu (Jayatillleke et al 2013) reported pesticide levels in urine samples from CKDu patients and from non-CKDU controls. The CKDu subjects with pesticide levels above thresholds were: 2,4-D (3.5%), Pentachlorophenol (1.2%), Chlorpyrifos (10.5%), Parathion (0%), Carbaryl (10.5%), Napthalene (10.5%) and Glyphosate (3.5%). Thus, except for Chlorpyrifos, Carbary and Napthalene, the other pesticides like glyphosate fall in the experimentally uncertain grey area of "less than 4%".

The (paddy) farmers in the CKDu-endemic areas in Sri Lanka use relatively less pesticides and agrochemicals than in the tea and vegetable plantations in the hill country. The hill country is the source of major rivers in Sri Lanka. Dharma-wardana et al (2015) considered the non-point source transport of phosphate by the irrigation waters of one of the major rivers (Mahaweli) of Sri Lanka, but the amounts of Cd and other metal toxins and pesticides transported are quite negligible, being present in parts per million compared to macro-nutrients. Diyabalanage et al (2016) confirmed that metal-toxin levels in the Mahaweli River are indeed below maximum allowed limits (MALs). Similarly, (Jayasinha et al 2015) showed that toxin levels in irrigation waters were below the MALs and hence required no reverse-osmosis treatment to render them safe. McLaughlin et al (1996) in their Review also disregard irrigation-water inputs of Cd into farm soils in Australia, and those arguments apply equally well to pesticide runoff. Aravinna et al (2017) found that the translocation of pesticides by irrigation waters of the Mahaweli River was negligible.

Although the yearly average for pesticides levels can be very low when two monsoons act to "wash out" the environment, short-time concentrations can spike. The presence of time-varying amounts of diazinon, propanil etc., in aquatic bodies has been reported in many studies. Jayasiri et al (2022) found that pesticide detection occurs in sporadic peaks linked to their application times. Given the short persistence lifetimes of about a day or less for diazinon and propanil at 30-36 degrees Celsius prevalent in the regions studied, these spikes of pesticide-residues rapidly drop to zero, and fully clear after monsoonal events.

Vlahos et al. (2022) find higher levels of insecticides (e.g., diazinon, DDE, propanil, endosulfan) in a "single-time" study of Wilgamuwa, Sri Lanka. They do not report persistence lifetimes. These toxins cause glomerular damage and liver damage (e.g., for diazinon, see Cakici et al 2013), while CKDu shows tubular-interstitial damage (Nanayakkara 2012) without liver damage. Hence these toxins *cannot* be linked to CKDu without considerable justification. A driving force behind the conclusions of Vlahos seems to be the *assumed* negative effects of the Green Revolution (GR). That GR sharply increased food production, eliminated malnutrition, and nearly doubled life expectancy, strengthening general health must also be considered while taking toll of possible increased environmental pollution.

Vlahos et al conclude "that agrochemical use in paddy and other agricultural practices … of the Green Revolution in Sri Lanka may now be contributing to ill health, rapid progression of disease, and mortality". They propose "reducing … agrochemical contaminants in Sri Lanka and other tropical countries to reduce … CKDu. These conclusions, based on a "single time-point analysis", tantamount to establishing an etiology of CKDu is unsupported even by the evidence presented by Vlahos et al. They do not satisfy even the simplest of Bradford-Hill criteria for causation. In particular, (i) similar but non-persistent pesticide excesses have been detected sporadically in most parts of the country including where there is no CKDu; (ii) the pesticides detected in Vlahos et al cause both hepatotoxicity and glomerular damage while CKDu is associated with tubulo- interstitial damage where no hepatotoxic symptoms have been reported; (iii) the pesticides detected have short half-lives and are used over short periods during farming; so the one-time measurement is misleading; (iv) farming communities that use pesticides in the same way but remain essentially without CKDu are found to exist adjacent to those with CKDu, as also seen in the Ginnoruwa area which is within 15-20 km of the area studied by Vlahos et al. (v) the CKDu prevalence seems to correlate with local geomorphology rather than with farming, as discussed in a previous section. However, associations do not imply causative action except that the work of Wasana et al (2017) and Thammitiyagoda et al (2017) provide a framework for possible causes.

**Consequence of dubious etiological claims causing public fear**.
During the year 2011 (and possibly earlier) Dr. Nalin De Silva, Dean of the faculty of Science of the University of Kelaniya, and Dr. Channa Jayasumana, a post-graduate student working on CKDu made claims stating that CKDu is caused by the presence of arsenic in the endemic areas. This claim was "supported" by invoking a clairvoyant, Ms. Priyani Senanayake who is alleged to communicate with divine beings like *God Natha* (DeSilva 2013), recognized both in Buddhist and Hindu religious lore.

Experimental work was done at the Kelaniya University with a view to confirm the revelation, and to satisfy "western science". Athureliye Ratana Thera (a Buddhist prelate as well as member of parliament) was also associated with these claims. However, De Silva writes "while we are grateful to the Thera for help,... we had finished our preliminary work in our laboratories before the Thera went to Malaysia [to obtain confirmation of the presence of As from accredited laboratories]. Our announcement to the media was made in January 2011 while the Thera went to Malaysia after March 2011".

Although several independent studies (Nanayakkara 2012, Jayatilleke et al 2013) had already shown that arsenic concentrations in the soil, water table and the environment of the affected areas were very low, a politicized campaign claiming that any arsenic (and heavy metals like Cd, Pb) found in the soil must have come via imported fertilizers and pesticides was launched. The claim that CKDu was caused by agrochemicals got a boost when a multi-level hypothesis was presented in 2014 that arsenic in the water of the endemic areas somehow replaces the phosphorus atoms in glyphosate molecules to form a hither-to unknown arsenic analogue of the glyphosate molecule. This postulated arseno-glyphosate was further postulated to chelate with calcium ions from the hard water of the endemic areas and form a hypothesized toxic compound causing CKDu.

Elementary chemistry and a table of bond energies should have been enough to completely dismiss every aspect of this scaffold of "hypotheses" adduced for the etiology of CKDu . This overall hypothesis, advanced by C. Jayasumana, S. Gunatillake and the clairvoyant contributor P. Senanayake submitted to the "Int. J. of Environ. Res. Public Health" on February 11th 2014, appeared in print on the 20th of February, i.e., *within nine days*. This so-called "journal" was in Biel's list of predatory journals (Brezgov 2019, Bohannon 2013) that publish without peer review, purely on payment of page charges. The publication immediately found support and world-wide publicity especially from groups carrying out a global campaign against glyphosate because of its importance as a herbicide in growing genetically modified crops (Ritterman 2014).

The hypothesis soon became a "fact" accepted by the general public as the "cause" of CKDu. Sri Lanka banned the use of glyphosate in 2014. The crippling effect on the tea industry, devastation of staples like maize and other crops (Marambe et al 2020), as well as the influx of smuggled glyphosate led to some relaxation of the glyphosate ban since 2018. The campaign was linked with a cry for traditional "organic agriculture". Influential members of the government medical officer's association (GMOA) strongly supported the ban, citing agrochemicals as the cause of CKDu as well as non-communicable diseases. Prominent medical men cited the "precautionary principle" to justify a ban even without clear evidence. Although these were countered by other scientists, legislators were more influenced by those who cried wolf and warned against agrochemicals.

Claims of bountiful harvests in ancient times when using traditional agriculture became common currency. The demonization of agrochemicals, and the stigmatization of critics as "paid agents of agrochemical companies" became popular. In April 2021 the Sri Lankan President banned the use of all agrochemicals and declared a "100% organic" policy. European eco-activists at the Glasgow climate summit received Rajapaksa's announcement with accolades.

The activist monk Ven. Ratana, and Dr. Jayasumana, lead author of the Arsenic+glyphosate+hardwater etiology of CKDu were powerful figures in the new government. Even the American Association for advancement of science ended up making wrong decisions in awarding a prize (Bodnar 2019) to Dr. Channa Jayasumana and Dr. Sanath Gunatilleke who "battled corporate interests while determining the cause of a kidney disease epidemic that claimed tens of thousands of lives"! Why was the co-author Ms. Seananayake, the clairvoyant lady, excluded from the AAAS prize?

The naive belief that the government could save foreign exchange needed to buy agrochemicals, and that all the manure and "natural pesticides" needed could be produce locally in the country had also played a part. The political leaders including ministers like Dr. Jayasumana may have not realized that organic manure is very low in N, P, etc., and that several tonnes of organic manure are needed to replace a kg of mineral fertilizer. Officials and scientists who predicted a loss of harvest lost their jobs. The dire consequences of what began as "divine-inspired" pseudoscience on the etiology of CKDu will take decades to play out its virulence.

**Postscript.**
The banning of fertilizers and agrochemicals since April 2021 led to a sharp decline in agricultural export revenue and disruption in the plantation sector. The attempts to supply organic fertilizers (compost manure, bone meal etc.) collapsed due to the sheer volume of the needed fertilizers. If 100 kg of mineral fertilizers were need for a few hectares of crop, it translates into several metric tonnes of organic fertilizer. While urea in any form (normal or slow release) may contain up to 46% nitrogen, organic fertilizers rarely reach even 2% nitrogen content.

Making one tonne of compost needs 4-10 tonnes of inputs of unprocessed organic matter. Such large inputs are simply not available, and even when available in sufficiently clean form, require expensive fuel and machinery to move such organic matter like farm refuse, acceptable urban garbage, aquatic weeds etc., pulled out and assembled for the purpose of composting.

Most compost pits and facilities are unevenly aerated, inhomogeneous and produce large amounts of methane. Methane is a green-house gas (GHG) which is 30 times more potent and more objectionable than carbon dioxide (the standard GHG). Furthermore, aeration of composting systems to encourage aerobic fermentation leads to increased emissions of nitrogen oxides; these are 300 times more potent GHGs as compared to carbon dioxide.

Hence increased organic farming to fully meet basic levels of food security is not only impossible due to lack of raw materials and processing costs, but it also creates a very serious a climate threat. In any case, Sri Lanka soon found that it could not produce even a tenth of the needed fertilizers, even with the army conscripted to produce compost.

Nevertheless, some eco-extremists and their apologists incorrectly claim that the only fault made by Sri Lanka was that it tried to "go too fast, too soon". Eco-activists underplay the impossibility of scaling up organic agriculture that currently produces a mere 2% of the world food supply. Other organic-farming analysts claim that it could be done, but require imposing vegetarianism on a world population reduced to half its current value. Furthermore, these analysts continue to ignore the negative effects of increased GHG emissions and increased need for land and water when switching over to organic farming, thus threatening biodiversity. The European *Green Deal*, as well as resolutions at the World Economic Forum at Davos show that many world leaders are misled by the rhetoric of fear-mongering and eco-extremist promises (Dharma-wardana 2023) of a "toxin-free" world.

Once the move to make organic fertilizer locally failed, Sri Lanka attempted to import organic fertilizers from China. Organic fertilizers contain living organisms that are consistent with the ecology of the location where the manure was made, but they cannot be safely transported to other environments and climate zones. Quarantine laws apply. Yet, Sri Lanka was so desperate that it was ready to even waive quarantine requirements.

However, analyses by Sri Lankan scientists showed that the imported organic fertilizer had unacceptable organisms like Erwiana (Daily Mirror, 2021), and hence the fertilizer had to be sent back without even unloading onto Sri Lanka's shores, even though Sri Lanka had paid out foreign exchange (in short supply) for these unwise purchases. In desperation, Sri Lanka spent further foreign exchange to import nano-urea from India, while still claiming that it is sticking to its "organic farming" targets, and falsely publicizing that the Indian liquid nano-urea is "organic"!

In spite of such last-minute efforts the farmers faced strongly diminished harvests, and when the next planting season (in 2022) arrived, they revolted due to lack of fertilizers; they burnt effigies of the minster of agriculture and other political figures. Instead of planting crops, they took to the streets (Dharma-wardana 2022b) and revolted.

This initial farmer's revolt was quickly taken over by other disgruntled urban groups who were affected by shortages of fuel and liquified butane gas used for cooking. These shortages were caused by the fall in foreign-exchange income that arose from the collapse of the agriculture sector, collapse of tourism and trade due to the Covid-19 pandemic, as well as the impact of the Ukrainian war. The president of the country had to flee the country on the 9$^{th}$ July 2022 to escape the wrath of the popular uprising (known locally as "aragalaya") that occurred (Ridley 2022), fortunately without significant bloodshed. It brought in a new care-taker government in July 2022.

**References.**